\documentclass[preprint1]{aastex}

\newcommand{\be}{\begin{equation}}
\newcommand{\ee}{\end{equation}} 

\shorttitle{CMB Large-scale Structure}
\shortauthors{Melia}

\begin{document}

\title{Cosmological Implications of the CMB Large-scale Structure}
\author{Fulvio Melia\thanks{John Woodruff Simpson Fellow.}\\
Department of Physics, The Applied Math Program, and Department of Astronomy,\\  
The University of Arizona, AZ 85721, USA\\
\email{fmelia@email.arizona.edu}}

\begin{abstract}
The Wilkinson Microwave Anisotropy Probe (WMAP) and {\it Planck}  may have
uncovered several anomalies in the full cosmic microwave background (CMB)
sky that could indicate possible new physics driving the growth of
density fluctuations in the early Universe. These include an unusually
low power at the largest scales and an apparent alignment of the
quadrupole and octopole moments. In a $\Lambda$CDM model where the
CMB is described by a Gaussian Random Field, the quadrupole and octopole
moments should be statistically independent. The emergence of these
low probability features may simply be due to posterior selections from
many such possible effects, whose occurrence would therefore not be as unlikely
as one might naively infer. If this is not the case, however, and if these 
features are not due to effects such as foreground contamination, their 
combined statistical significance would be equal to the product of their 
individual significances. In the absence of such extraneous factors,
and ignoring the biasing due to posterior selection, the missing large-angle 
correlations would have a probability as low as $\sim 0.1\%$ and the
low-$l$ multipole alignment would be unlikely at the $\sim 4.9\%$
level; under the least favourable conditions, their simultaneous observation 
in the context of the standard model could then be likely at only
the $\sim 0.005\%$ level. In this paper, we explore the possibility
that these features are indeed anomalous, and show that the 
corresponding probability of CMB multipole alignment in the 
$R_{\rm h}=ct$ Universe would then be $\sim 7-10\%$, 
depending on the number of large-scale Sachs-Wolfe
induced fluctuations. Since the low power at the largest spatial
scales is reproduced in this cosmology without the need to invoke
cosmic variance, the overall likelihood of observing both of these
features in the CMB is $\geq 7\%$, much more likely than in
$\Lambda$CDM, if the anomalies are real. The key physical ingredient
responsible for this difference is the existence in the former
of a maximum fluctuation size at the time of recombination,
which is absent in the latter because of inflation.
\end{abstract}


\section{Introduction}
The Wilkinson Microwave Anisotropy Probe (WMAP) and the {\it Planck} satellite
have revolutionized our ability to study anisotropies in the cosmic microwave
background (CMB) with a precision that is now permitting us to examine the
structure of the Universe on all scales (Bennett et al. 2003; Ade et al. 2013).
But several apparent anomalies may be indicating possible new physics driving the
origin of density fluctuations in the early Universe and their evolution
into the large-scale structure we see today. These peculiarities include an
unusually low power at the largest scales (Spergel et al. 2003), as well as an
apparent alignment of the quadrupole and octopole moments (Tegmark et al.
2003; de Oliveira-Costa et al. 2004; Hansen et al. 2004; Eriksen et al. 2004;
Schwarz et al. 2004; Land \& Magueijo 2005). Prior to the {\it Planck} era,
these features had variously been attributed to astrophysical, instrumental,
or cosmological causes, and even faulty data analysis or {\it a posteriori}
statistics (Copi et al. 2009). The possibility that these features might
be due to instrumental effects, however, has recently been weakened by the
{\it Planck} mission, which has confirmed the low power on the
largest scales and an alignment between $9^\circ$ and $13^\circ$ of the
quadrupole and octopole orientations (Ade et al. 2013). The low power and
alignment are puzzling because the probability of either occurring within
the context of the standard model ($\Lambda$CDM), under the
assumption that neither is the result of mere posterior selection biases
(Bennett et al. 2011), is less than $\sim1\%$; the chance to measure 
the sky with both has been quantified at $<10^{-6}$ (Sarkar et al. 2011).

The power on large scales is characterized  in terms of the angular
correlation function $C(\theta)$ (defined in Eq.~2 below). According to the
observations, $C(\theta)$ essentially vanishes at angular separations greater
than about $60^\circ$ (Spergel et al. 2003; Ade et al. 2013), confirming what
was first  measured a decade earlier with the Cosmic Background Explorer
(COBE; Hinshaw et al. 1996; Wright et al. 1996). The absence of any angular
correlation at the largest scales could be a problem for the standard
model because it disagrees with the expectations of an inflationary scenario
(Guth 1981; Linde 1982). Yet without inflation, $\Lambda$CDM could not account
for the apparent uniformity of the CMB (other than the aforementioned
anisotropies at the level of 1 part in 100,000) across the sky.

But though the reality of these anomalies is no longer questioned, the
possible reasons for their existence, and their significance, are still being
discussed and evaluated. As Copi et al. (2010) have pointed out in their
review on this subject, four classes of explanations have thus far been
proposed, including astrophysical causes, such as foreground effects,
faulty data analysis, instrumental systematics, and perhaps cosmological
reasons. One can find faults with each of these explanations, so no
consensus has yet been reached on which is the most likely to account
for the observations. Nonetheless, an obvious possible cause of the
anisotropy is contamination by a pernicious foreground (see, e.g.,
Slosar and Seljak 2004; Bielewicz et al. 2005; Copi et al. 2006).
In this class of explanations, some workers have suggested that the
observed quadrupole-octopole alignment might be due to the Rees-Sciama
effect (Rakic et al. 2006; Rakic and Schwarz 2007), interstellar dust
(Frisch 2005), the presence of local voids (Inoue and Silk 2006), or
even the Sunyaev-Zeldovich effect (Peiris and Smith 2010). Another
proposal by Vale (2005) argues that the unexpected anisotropy might
be due to a moving lens effect associated with the Great Attractor,
though its influence may be too small to fully account for the
observations (Cooray and Seto 2005). And on a more local scale,
Dikarev et al. (2008,2009) have argued that the solar system dust
could give rise to sizable levels of microwave emission and absorption.

Artifacts produced by faulty data analysis could also produce
unexpected anisotropies. When using reconstructed full-sky maps
(e.g., Bennett et al. 2003; Tegmark et al. 2003; Eriksen et al. 2004),
a sky cut of just a few degrees produces errors in the reconstructed
anisotropy pattern and the directions of multipole vectors that are
too large to allow quantitative conclusions concerning the apparent
alignments (Copi et al. 2004). Efstathiou et al. (2010) have argued
that one should instead use maximum likelihood estimators to the
cut-sky maps to reliably reconstruct the CMB anisotropy distribution.
They argue that estimating the probability of seeing these anomalies
based on the pixel approach using the cut sky is simply a
coincidence, and that a more reliable result is produced from their
maximum-likelihood reconstruction technique.

Prior to the observations by {\it Planck}, which largely
confirmed the reality of the CMB anomalies, instrumental explanations
also played a role. These are perhaps not as likely now that WMAP and
{\it Planck} confirm each other's measurements, though one can not be
sure that all systematics have been eliminated (Bennett et al. 2003).
For example, imperfections in the instrument may couple with dominant
signals from the sky to create anomalies (Gordon et al. 2005).

The WMAP team itself considers the CMB anomalies, such as the
low-multipole alignment, to be real, though they question the significance
of these results and the possibility that they may be due to cosmological
influences (Bennett et al. 2011). They note that Chiang et al. (2007)
find that the lowest spherical harmonic modes in the ILC map are
significantly contaminated by foregrounds. And Park et al. (2007)
conclude that the residual foreground emission based on their own
analysis is not statistically important to the large-scale modes of
the CMB. Bennet et al. (2011) also support the view that instead
of early universe effects, the apparent low-multipole alignment
may be due to the integrated Sachs-Wolfe (ISW) effect associated the
local ($z<0.3$) density fields. Francis and Peacock (2010) estimated
this density field from the 2MASS and SuperCOSMOS galaxy catalogs
and used these to calculated the ISW effect within this volume.
When they removed their estimated ISW contribution from the WMAP
map, the quadrupole amplitude increased while that of the octopole
remained relatively unchanged. More importantely, they concluded
that no significant quadrupole-octopole alignment remained after this
subtraction.

One therefore must be wary about placing too much confidence in the notion
that the low-multipole alignment is necessarily a signature of cosmological
effects in the early Universe. However, the origin of this alignment,
which is generally considered to be real, is still unknown. In this paper,
we consider whether a cosmological basis for this low-$l$ multipole alignment
could be used to distinguish between the $\Lambda$CDM and $R_{\rm h}=ct$
expansion scenarios. Specifically, we seek to address the question of
whether an alignment that may be statistically unexpected in the context
of the standard model could instead be more in line with that expected
in the $R_{\rm h}=ct$ Universe (Melia 2007; Melia \& Shevchuk
2012; Melia 2013a) which, unlike $\Lambda$CDM, did not undergo a period
of early inflation. Earlier, we showed that the absence of power on large
scales exhibited by the angular correlation function might be evidence
in support of this model simply because it does not require inflation
(Melia 2014). Here, we discuss whether an absence of early inflation
might also provide an explanation for the apparent alignment of the
CMB's quadrupole and octopole moments. We will consider the 
conditions under which large-scale fluctuations in the $R_{\rm h}=ct$
cosmology could account for the low-$l$ portion of the spectrum. As 
we shall see, the constraints emerging from this analysis will allow us 
to determine the kinds of all-sky map one would associate with the 
$R_{\rm h}=ct$ condition. We will carry out a statistical analysis of 
thousands of simulated renderings to calculate the probability of 
seeing an apparent alignment of the CMB quadrupole and octopole 
moments in the $R_{\rm h}=ct$ Universe, and compare this with 
that expected in $\Lambda$CDM.

Since this is the first attempt at simulating the fluctuation spectrum
in $R_{\rm h}=ct$, we will necessarily restrict our attention to the
influence most responsible for producing the quadrupole and octopole
moments---the Sachs-Wolfe effect (Sachs \& Wolfe 1967). Other phenomena,
such as the Baryon Acoustic Oscillations, influence primarily the
fluctuation growth on sub-degree scales. We will incorporate these
effects in future applications of this work, but for the sake of
keeping the interpretation of the results as simple as possible,
focusing primarily on the largest fluctuations, we will not include
them here.

\section{The CMB Angular Power Spectrum and Correlation Function}
The CMB temperature anisotropies $\Delta T(\Omega)/T$ extracted from
WMAP and {\it Planck} may be written as an expansion using spherical
harmonics $Y_{lm}(\hat{\bf n})$,
\begin{equation}
{\Delta T(\Omega)\over T}=\sum_{lm}a_{lm}Y_{lm}\;,
\end{equation}
from which one can then determine the two-point angular
correlation function (for directions $\hat{\bf n}_1$ and $\hat{\bf n}_2$):
\begin{equation}
C(\cos\theta)\equiv \langle T(\hat{\bf n}_1)T(\hat{\bf n}_2)\rangle=
{1\over 4\pi}\sum_l (2l+1)C_l P_l(\cos\theta)\;.
\end{equation}
Statistical independence implies that
\begin{equation}
\langle a^*_{lm}a_{l'm'}\rangle\propto\delta_{ll'}\,\delta{mm'}\;,
\end{equation}
and statistical isotropy further requires that the constant of proportionality
depend only on $l$, not $m$:
\begin{equation}
\langle a^*_{lm}a_{l'm'}\rangle=C_l\,\delta_{ll'}\,\delta{mm'}\;.
\end{equation}
The constant
\begin{equation}
C_l={1\over 2l+1}\sum_m |a_{lm}|^2
\end{equation}
is the angular power of the multipole $l$.

A comparison of the function $C(\theta)$ predicted by the $R_{\rm h}=ct$
Universe with that observed by WMAP was the primary goal of our previous
paper (Melia 2014). Our focus here will be the second CMB anomaly discussed
above, i.e., the apparent alignment of $C_2$ and $C_3$. To quantify the statistical
significance of this alignment, we will follow the procedure developed by
(de Oliveira-Costa et al. 2004). Other techniques have been utilized since
then (Copi et al. 2009), but they all appear to confirm each other's
results, so for now, at least, we will base our assessment on the former
approach.

The method treats the CMB map as a wave function,
\begin{equation}
{\Delta T\over T}(\hat{\bf n})\equiv \psi(\hat{\bf n})\,,
\end{equation}
and seeks to find the axis $\hat{\bf n}$ about which the ``angular
momentum" dispersion
\begin{equation}
\langle\psi |(\hat{\bf n}\cdot{\bf L})^2|\psi\rangle=\sum_m
m^2|a_{lm}(\hat{\bf n})|^2
\end{equation}
is maximized. The coefficients $a_{lm}(\hat{\bf n})$ correspond
to the spherical harmonics in a rotated coordinate system with the
$z$-axis in the $\hat{\bf n}$ direction. For the actual CMB map,
the maximization is performed by evaluating Eq.~(7) at all
the unit vectors $\hat{\bf n}$ corresponding to the pixel centers.
We will follow essentially the same approach, first producing a
rendering of the large-scale fluctuations on the whole sky, and
then maximizing the angular momentum dispersion using the same
equation (more on this below). Our synthetic images each contain
$180\times 360$ pixels in $\theta$ and $\phi$, respectively.

Previous papers have reported the preferred axes
$\hat{\bf n}_2$ and $\hat{\bf n}_3$ for the quadrupole
and octopole moments to be
\begin{eqnarray}
\hat{\bf n}_2&\sim&(-0.1145,-0.5265,0.8424)\;,\cr
\hat{\bf n}_3&\sim&(-0.2578,-0.4207,0.8698)\;,
\end{eqnarray}
respectively, i.e., both roughly in the direction $(l,b)\sim
(-110^\circ,60^\circ)$ in Virgo (de Oliveira et al. 2004). In $\Lambda$CDM,
a crucial ingredient is cosmological inflation---a brief phase of
very rapid expansion from approximately $10^{-35}$ seconds
to $10^{-32}$ seconds following the big bang, forcing the Universe
to expand much more rapidly than would otherwise have been feasible
solely under the influence of matter, radiation, and dark energy. This
accelerated expansion would have driven the growth of fluctuations
on all scales, resulting in an angular correlation at all angles (which
does not appear to be consistent with the WMAP and {\it Planck} results;
Melia 2014). Therefore, in $\Lambda$CDM, the unit vectors $\hat{\bf n}_2$
and $\hat{\bf n}_3$ should be independently drawn from a
distribution in which all directions are equally likely. This means
that the dot product $\hat{\bf n}_2\cdot\hat{\bf n}_3$ should
be a uniformly distributed random variable on the interval $(-1,1)$.
But as is well known, Eq.~(7) does not distinguish between
$\hat{\bf n}$ and $-\hat{\bf n}$, so the maximization procedure
finds a preferred axis, not a preferred direction. The alignment
should therefore be quantified on the basis of $|\hat{\bf n}_2\cdot
\hat{\bf n}_3|$, which instead has a uniform distribution on the
interval $(0,1)$ (de Oliveira-Costa et al. 2004).

The anomaly emerges when we determine from Eq.~(8) that the
observed value of this dot product is $|\hat{\bf n}_2\cdot
\hat{\bf n}_3|\approx 0.9838$, corresponding to a separation
of only $10.3^\circ$. An alignment this good happens by chance
only once in 62 realizations, suggesting that the probability
of finding a random octopole axis within a circle of radius
$10.3^\circ$ of the quadrupole axis should be less than a few
percent. Within the context of $\Lambda$CDM, this alignment is
therefore a statistically significant anomaly, in terms of
the standard definition in which an outcome with probability $<5\%$
is considered to be significantly anomalous. (Note, however, that
this may still be $<3\sigma$.)

Recognizing that producing an accurate map of the WMAP
data depends critically on correctly identifying the background
(or more accurately in this case, the foreground),
this calculation has been repeated on several occasions, with an
ever increasing precision of the foreground subtraction. The numbers
themselves have changed somewhat, but all subsequent
measurements have confirmed the early conclusions. The
most likely outcome currently appears to be an alignment
angle $3.8^\circ<\theta_{23}<18.2^\circ$ (Park et al. 2007).
Even with such a broadened uncertainty range, however, an
alignment within $\theta_{23}\sim 18^\circ$ should occur only
$\sim 4.9\%$ of the time, making it a marginally statistically
significant anomaly within the standard model. And as
we noted earlier, the latest results from the Planck mission
appear to confirm these earlier conclusions, though with a
somewhat different range of possible angles, $\sim
9^\circ - 13^\circ$. The implied alignment
for both WMAP and Planck sits comfortably within the
interval $\theta_{23}\le 18^\circ$, and we will therefore
use this conservatively large range for all of the analysis
that follows.

\section{Large-Scale Fluctuations in the $R_{\rm h}=ct$ Universe}
The $R_{\rm h}=ct$ Universel is an FRW cosmology (Melia 2007; Melia 
\& Shevchuk 2012) that adheres very closely to the restrictions imposed
on the theory by the Cosmological Principle and the Weyl postulate (Weyl
1923). Taken seriously, these two basic tenets force the gravitational
horizon $R_{\rm h}$ (recognized more commonly as the Hubble radius)
to always equal $ct$, where $t$ is the cosmic time. It is easy to
convince oneself that this equality forces the expansion rate
to be constant, so the expansion factor $a(t)$ appearing in the
Friedmann equations  must be $t/t_0$ (utilizing the convention that
$a(t_0)=1$ today), where $t_0$ is the current age of the Universe.

$\Lambda$CDM appears to be an approximation to this cosmology
because it adopts a specific set of constituents for the energy density
$\rho$, though without the important constraint that the overall equation
of state must be $p=-\rho/3$, where $p$ and $\rho$ are the total
pressure and density, respectively. Therefore, $R_{\rm h}$ in $\Lambda$CDM
fluctuates about the mean it would otherwise always have, leading
to the awkward situation in which the value of $R_{\rm h}(t_0)$
is equal to $ct_0$ today, but in order to achieve this ``coincidence",
the Universe had to decelerate early on, followed by a more recent
acceleration that exactly balanced out the effects of its slowing
down at the beginning. It is specifically this early deceleration
in $\Lambda$CDM that brings it into conflict with the near uniformity
of the CMB data, requiring the introduction of an inflationary phase
to rescue it. As shown in Melia (2014), however, the recent assessment
of the observed angular correlation function suggests that a
possible reason for the lack of correlation at large angles may be
the absence of an inflationary episode.

Over the past several years, we have tested the predictions of
$R_{\rm h}=ct$ against several different types of observational
data, both at low and high redshifts. These efforts have demonstrated
that, statistically speaking, the $R_{\rm h}=ct$ Universe is more
likely than $\Lambda$CDM to be the correct description of nature
(see, e.g., Melia \& Maier 2013; Melia 2013b; Wei et al. 2013).
In this paper, we add to the comparative study of $\Lambda$CDM
versus $R_{\rm h}=ct$, by examining how the large-scale fluctuations
in these two cosmologies account for the CMB data, focusing on the
question of whether the near alignment of the quadrupole and octopole
moments ought to be viewed as statistically significant.

As shown in Melia \& Shevchuk (2012) and Melia (2014), density
fluctuations $\delta\equiv\delta\rho/\rho$, written as a wavelike decomposition
\begin{equation}
\delta=\sum_{\kappa}\delta_{\kappa}(t)e^{i\vec\kappa\cdot{\bf r}}\;,
\end{equation}
satisfy the differential equation
\begin{equation}
\ddot{\delta}_\kappa+{3\over t}\dot{\delta}_\kappa={1\over 3}c^2\left({\kappa\over a}\right)^2\delta_\kappa\;.
\end{equation}
The way perturbation growth is handled in $R_{\rm h}=ct$, leading to Equation~(10),
 is somewhat different from $\Lambda$CDM, so let's take a moment to briefly
describe the origin of this expression. The chief difference between
$\Lambda$CDM and $R_{\rm h}=ct$ is that one
must guess the constituents of $\rho$ in the former, assign an individual
equation of state to each, and then solve the growth equation derived for
each of these components separately. This is how one handles a situation
in which the various species do not necessarily feel each other's pressure,
though they do feel the gravitational influence from the total density.
The coupled equations of growth for the various components can be quite
complex, so one typically approximates the equations by expressions that
highlight the dominant species in any given era. For example, before
recombination, the baryon and photon components must be treated
as a single fluid, since they are coupled by frequent interactions in
an optically-thick environment. During this period, $\Lambda$CDM
assumes that ``dark energy" is smooth on scales corresponding
to the fluctuation growth, and treats the baryon-photon fluid
as a single perturbed entity with the pressure of radiation and
an overall energy density corresponding to their sum.
Once the radiation decouples from the luminous
matter, all four constituents (including dark matter) must be
handled separately.

The situation in $R_{\rm h}=ct$ is quite different for several
reasons. First of all, the overall equation of state in this cosmology
is not forced on the system by the constituents; it is the other way
around. The symmetries implied by the Cosmological Principle and
Weyl's postulate together, through the application of general
relativity, only permit a constant expansion rate, which means
that $p = -\rho/3$.  The expansion rate depends on the total
energy density, but not on the partitioning among the various
constituents. Instead, the constituents must partition themselves
in such a way as to always guarantee that this overall equation
of state is maintained during the expansion.

And since the pressure is therefore a non-negligible fraction of $\rho$
at all times, one cannot use the equations of growth derived
from Newtonian theory (commonly employed in $\Lambda$CDM),
since $p$ itself acts a source of curvature. One must therefore
necessarily start with the relativistic growth equation (numbered
41 in Melia \& Shevchuk 2012), which correctly incorporates all of the
contributions from $\rho$ and $p$. This equation is ultimately derived
from Einstein's field equations using the perfect-fluid form of the
stress-energy tensor, written in terms of the total $\rho$ and total
$p$, but without specifying the sub-partitioning of the density
and pressure among the various constituents. With this approach,
there is only one growth equation.

On occasion, it is also necessary to use the relativistic growth
equation in $\Lambda$CDM. But there, one typically chooses a
regime where a single constituent is dominant, say during the
matter-dominated era, and then one assumes that $\rho$ is
essentially just the density due to matter (for which also
$p\approx 0$). But in general, since the pressure appearing
in the stress-energy tensor is the total pressure, one cannot
mix and match different components that may or may not ``feel"
each other's influence (as described above). So in fact using
the relativistically correct growth equation is difficult in $\Lambda$CDM,
unless one can make suitable approximations in a given regime.

In $R_{\rm h}=ct$, on the other hand, the total pressure is
always $-\rho/3$, so the key question is whether all of the
constituents participate in the perturbation growth, or
whether only some of them do. There is no doubt that the
baryons and photons are coupled prior to recombination. In
$\Lambda$CDM, one assumes that dark energy is coupled
only weakly, acting as a smooth background. In $R_{\rm h}=ct$, dark energy
cannot be a cosmological constant. One therefore assumes that
during the early fluctuation growth, everything is coupled strongly
in order to maintain the required total pressure $-\rho/3$.
This may change locally once the matter has clumped
if it decouples from dark energy on such small scales.

In short, there is one assumption made in each cosmology. In $\Lambda$CDM,
dark energy is a cosmological constant that remains smooth while
the baryon-photon fluid is perturbed at early times. In $R_{\rm h}=ct$,
dark energy cannot be a cosmological constant, and everything is
coupled strongly at early times, so the perturbation affects the
total energy density $\rho$. One must always use the correct
relativistic growth equation, which includes $p$ as a source
of gravity.

In the end, this equation simplifies considerably because the active mass
in $R_{\rm h}=ct$  is proportional to $\rho+3p=0$, and therefore the gravitational term
normally appearing in the standard model is absent. But this does not mean that $\delta_\kappa$
cannot grow. Instead, because $p<0$, the (usually dissipative) pressure term on the right-hand-side
here becomes an agent of growth. Moreover, there is no Jeans length scale. In its place is the
gravitational radius, which we can see most easily by recasting this differential equation in the form
\begin{equation}
\ddot{\delta}_\kappa+{3\over t}\dot{\delta}_\kappa-{1\over 3}{\Delta_\kappa^2\over t^2}\delta_\kappa=0\;,
\end{equation}
where
\begin{equation}
\Delta_\kappa\equiv {2\pi R_{\rm h}\over \lambda}\;.
\end{equation}
Note, in  particular, that both  the gravitational radius $R_{\rm h}$ and the fluctuation
scale $\lambda$ vary with $t$ in exactly the same way, so $\Delta_\kappa$ is therefore
a constant in time. But the growth rate of $\delta_\kappa$ depends critically on whether
$\lambda$ is less than or greater than $2\pi R_{\rm h}$.

A simple solution to equation~(11) is the power law
\begin{equation}
\delta_\kappa(t)=\delta_\kappa(0)t^\alpha\;,
\end{equation}
where
\begin{equation}
\alpha^2+2\alpha-{1\over 3}\Delta^2_\kappa=0\;.
\end{equation}
That is,
\begin{equation}
\alpha=-1\pm\sqrt{1+\Delta_\kappa^2/3}\;,
\end{equation}
so for small fluctuations ($\lambda<<2\pi R_{\rm h}$), the growing mode is
\begin{equation}
\delta_\kappa\sim \delta_\kappa(0)t^{\Delta_\kappa/\sqrt{3}}\;,
\end{equation}
whereas for large fluctuations ($\lambda>2\pi R_{\rm h}$), the dominant mode
\begin{equation}
\delta_\kappa\sim \delta_\kappa(0)
\end{equation}
does not even grow. The second mode decays away for both small and large fluctuations.

Insofar as the quadrupole and octopole moments are concerned, the most critical aspect of
the fluctuations implied by these equations is the maximum range over which they would
have grown. The required inflated expansion in $\Lambda$CDM drives the growth over
all scales. In the $R_{\rm h}=ct$ Universe, on the other hand, the growth is limited to a
maximum fluctuation size
\begin{equation}
\lambda_{\rm max}(t)\sim 2\pi R_{\rm h}(t)\;.
\end{equation}
Thus, since the comoving distance to the last scattering surface (at time $t_e$) is
\begin{equation}
 r_e=ct_0\int_{t_e}^{t_0}{dt'\over t'}=ct_0\ln\left({t_0\over t_e}\right)\;,
\end{equation}
the maximum angular size $\theta_{\rm max}$ of any fluctuation associated with the
CMB emitted at $t_e$ has to be
\begin{equation}
\theta_{\rm max}={\lambda_{\rm max}(t_e)\over R_e(t_e)}\;,
\end{equation}
where
\begin{equation}
R_e(t_e)=a(t_e)r_e=a(t_e)ct_0\ln\left({t_0\over t_e}\right)=ct_e\ln\left({t_0\over t_e}\right)
\end{equation}
is the proper distance to the last scattering surface at time $t_e$. That is,
\begin{equation}
\theta_{\rm max}\sim {2\pi\over \ln(t_0/t_e)}\;.
\end{equation}
For the sake of illustration, we note that the times $t_0=13.7$ Gyr and $t_e\approx 
380,000$ yrs from the standard model would imply  $\theta_{\rm max}\sim 34^\circ$.
It is the existence of this limit that allows the $R_{\rm h}=ct$ Universe to fit the
angular correlation function much better than $\Lambda$CDM, and we shall see
shortly that the existence of this limit also alters the probability of seeing a
low-multipole alignment of the CMB, rendering it statistically insignificant.

In the spirit of identifying the key elements of the theory responsible for the CMB
fluctuations, without necessarily getting lost in the details of the complex treatment
involving fluctuation growth on small and large scales, and the impact of transfer
functions that link the observed temperature variations to the incipient density
perturbations, we will here follow the same approach described in (Melia 2014),
which itself is based on simplified methods used in earlier applications (Efstathiou 1990).

The Sachs-Wolfe effect dominates the fluctuation growth on scales larger than 
$\sim 1^\circ$ (Sachs \& Wolfe 1967). In $\Lambda$CDM, the assumption is now made
that by the time these fluctuations have formed, one can ignore the contribution of
radiation pressure to the total active mass in the fluid, so that at these large 
wavelengths, the amplitude of the temperature fluctuation ought to scale solely with 
the local gravitational potential (Efstathiou 1990).  For this to be valid, one must
also assume that dark energy (presumably a cosmological constant) functions as
a smooth background. In $R_{\rm h}=ct$, the corresponding sequence of steps 
is similar, with analogous assumptions concerning the behavior of dark energy 
(which in this case cannot be a cosmological constant). In this model, the zero 
active mass condition, $\rho+3p=0$, applies on large scales, where the (Hubble)
flow is relatively smooth, but not necessarily on small scales once the matter has
clumped. At the beginning when the fluctuations start to grow, all the constituents, 
including matter, radiation, and dark energy, are coupled together and produce a 
total pressure $p=-\rho/3$. This is reflected in the form of Equation~(11) and its 
solutions. But as in $\Lambda$CDM, the assumption is made here that dark energy 
does not clump on scales comparable to matter. Thus, while matter still ``feels" 
the total pressure (due to radiation and dark energy), this pressure is relatively
uniform, and therefore does not contribute to the local potential.

We therefore assume that on this scale (as opposed to the larger, smooth Hubble 
flow), only the gravitational potential associated with the energy density $\rho$ 
influences the CMB. It is therefore not difficult to show that
\begin{equation}
{\Delta T\over T}\sim \delta\rho\,\lambda^2\;.
\end{equation}
The variance in density over a particular comoving scale $\lambda$ is given as
\begin{equation}
\left({\delta\rho\over \rho}\right)^2_\lambda\propto
\int_0^{\kappa\sim1/\lambda}P(\kappa')d^3\kappa'
\end{equation}
(see, e.g., Efstathiou 1990), where $P(\kappa)=\langle |\delta_\kappa |^2\rangle$
is the power spectrum. Not knowing the exact form of $P(\kappa)$ emerging from
the non-linear growth prior to recombination, we will follow the approach outlined
in Melia (2014) and parametrize it as follows,
\begin{equation}
P(\kappa)\propto \kappa-b\left({2\pi\over R_e(t_e)}\right)^2\kappa^{-1}\;,
\end{equation}
where the (unknown) constant $b$ is expected to be $\sim O(1)$.

This form of $P(\kappa)$ is based on the following reasoning.
One typically assumes a scale-free initial power-law spectrum, 
which is what one might have expected with or without the complex 
machinery of inflation. Such a spectrum is often referred to as a 
Harrison-Zeldovich-Peebles spectrum, since these were the
indivdiuals who first proposed it as appropriate for the initial 
conditions many years prior to the development of 
inflationary models. (In fact, a better signature of inflation is 
a small deviation from a scale-free spectrum and/or a 
mixture of tensor modes.) In $\Lambda$CDM, these
fluctuations grow and then expand on all scales during the required
inflationary phase. In $R_{\rm h}=ct$, the fluctuation growth is driven
by the (negative) pressure, represented by the term on the right-hand
side of equation~(10). Because there is no Jeans length,
fluctuations can in principle grow on all scales as well. However, this equation
also shows that what matters most is the ratio of the fluctuation length
$\lambda$ to the gravitational radius $R_{\rm h}(t)$ at time $t$.
The solution to this equation shows that only fluctuations with $\lambda 
< 2\pi R_{\rm h}$ will grow, and that those modes that grow, will grow rapidly,
given their strong dependence on $t$ (see equation~16).

One can see from this equation and the definition of $P(\kappa)$ that 
for $\lambda<<2\pi R_{\rm h}$, corresponding to large $\kappa$, we have
$\alpha\sim \kappa_0\kappa$, where $\kappa_0\equiv ct_0/\sqrt{3}$, so that
\begin{equation}
P(\kappa) \approx \kappa t^{2\kappa_0\kappa}\qquad({\rm large}\; \kappa)\;.
\end{equation}
On the other hand, for $\lambda>>2\pi R_{\rm h}$, corresponding to small
$\kappa$, we have $\alpha\sim (\kappa_0\kappa)^2/2$, and so
\begin{equation}
P(\kappa) \approx \kappa t^{(\kappa_0\kappa)^2}\qquad({\rm small}\; \kappa)\;.
\end{equation}
For a given value of the cosmic time (say at recombination, $t_{\rm e}$), we
can express this changing behavior for small and large $\kappa$ as a sum
of polynomials,
\begin{equation}
P(\kappa)=\sum_i c_i\kappa^i+\sum_j b_j\kappa^{-j}\;,
\end{equation}
with leading order terms $P(\kappa)\sim c_1\kappa+b_1\kappa^{-1}$. 
The choice of $c_1$ and $b_1$ shown in equation~(25) provides a reasonable
representation of the evolution with $\kappa$ from equation(26) to (27).

What this means physically is that the fluctuations will grow quickly in 
amplitude up to the size $2\pi R_{\rm h}(t)$, above which the growth is 
suppressed. The simple parametrization in equation~(25) incorporates these
essential effects: first, the initial seed spectrum is assumed to be
scale-free, which means that $P(\kappa)\sim\kappa$. Since the growth rate
depends critically on the ratio $R_{\rm h}/\lambda$, one would expect
$P(\kappa)$ to be dominated by the smaller wavelengths (i.e., the
larger $\kappa$'s), and be altered more and more for increasing
wavelengths (i.e., smallter $\kappa$'s). Since the growth rate 
decreases with decreasing $\kappa$, one would expect a greater 
and greater depletion in power. The second term in equation~(25) 
represents this effect.

Now, from equations~(24) and (25), we see that
\begin{equation}
\left({\delta\rho\over \rho}\right)^2_\lambda\propto
{1\over 4}\kappa^4-{1\over 2}b\left({2\pi\over R_e(t_e)}\right)^2\kappa^2\;.
\end{equation}
Defining the angle $\theta$ analogously with $\theta_{\rm max}$ in equation~(20),
we put
\begin{equation}
\theta\equiv{\lambda\over R_e(t_e)}\;,
\end{equation}
whereupon
\begin{equation}
\left({\delta\rho\over \rho}\right)^2_\lambda\propto
{1\over 4}\kappa^4\left(1-2b\theta^2\right)\;.
\end{equation}
Therefore, keeping only higher order terms in the binomial expansion
for the square root of the factor on the right, we find that
\begin{equation}
\delta\rho\sim {1\over\lambda^2}\left(1-b\theta^2\right)\;.
\end{equation}

Thus, the amplitude of the Sachs-Wolfe temperature fluctuations
follows the very simple form
\begin{equation}
{\Delta T\over T}\sim \left(1-b\theta^2\right)\;,
\end{equation}
but only up to the maximum angle $\theta_{\rm max}$ established earlier.

In comparing the angular correlation function $C(\theta)$ resulting from this
expression with that inferred from the WMAP data, we found earlier that the
general shape of $C(\theta)$ depends only weakly on the chosen values of
$b$ and $t_0/t_e$. Insofar as the large-scale fluctuations are concerned, therefore,
the principal feature of the $R_{\rm h}=ct$ Universe that distinguishes it from
$\Lambda$CDM is the existence of the maximum angle $\theta_{\rm max}$.
We anticipate that the outcome will be similar here if the two large-scale
anomalies are indeed linked in this cosmology.

\section{The CMB Power Spectrum for Low-$l$}
Throughout our discussion in this paper, we focus
solely on fluctuations induced by the Sachs-Wolfe effect,
ignoring other important physical processes, such as Baryon
Accoustic Oscillations, that dominate on scales of a few degrees,
or less. Since the values of $b$ and $t_0/t_e$ were essentially
identified from our study of the angular correlation function,
the principal unknown here is therefore the number $N_{SW}$ of
Sachs-Wolfe fluctuations across the sky. The CMB power spectrum
at angles $>>1^\circ$ arises from these, so it should be possible
to fit the data for $l<20$ using the $C_l$'s calculated from
Equations~(1)-(5) and (27), in order to infer the approximate
range of values of $N_{SW}$ implied by the observations.

  \begin{figure}[h]
   \centering
      \includegraphics[angle=0,width=8.3cm]{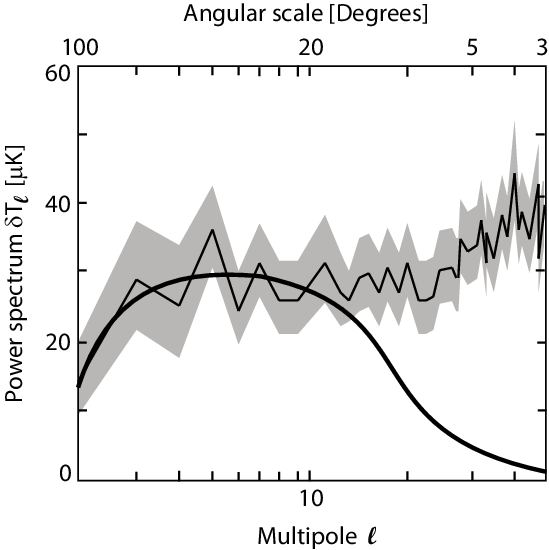}
      \caption{The theoretical CMB power spectrum due solely to
      Sachs-Wolfe-induced fluctuations in the $R_{\rm h }=ct$
      Universe (solid, thick curve), in comparison with the
      power spectrum measured from the full WMAP sky
      (thin, jagged line; Spergel et al. 2003; Tegmark et al. 2003).
      The gray region represents the one-$\sigma$ uncertainty. The
      number of fluctuations assumed for this simulation is
      $N_{SW}=5,000$. The power spectrum for $l> 20$ is
      dominated by small-scale physical effects, such as Baryon
      Acoustic Oscillations near the surface of last scattering, which
      are not included in our analysis. See text for other details.}
\end{figure}

The CMB power spectrum is calculated according to
\begin{equation}
\delta T_l^2=l(l+1)C_l/2\pi\;,
\end{equation}
with the angular power $C_l$ of multipole $l$ given in
Equation~(5). Several theoretical curves showing
$\delta T_l$ versus $l$ are shown in Figures~1--4,
for various choices of the 2 principal parameters at our disposal,
the constant $b$ in Equation~(27), and $N_{SW}$. The
CMB's angular correlation function also depends quantitatively
(though not qualitatively) on the ratio $t_0/t_e$, which defines
$\theta_{\rm max}$ in Equation~(22) (Melia 2014). Eventually,
detailed simulations of the fluctutations growth with the timeline
afforded by the $R_{\rm h}=ct$ Universe will provide a reliable
estimate of this ratio, so we won't have to treat it as an unknown
parameter. But for now, the results of the angular-correlation
function analysis indicate that the effects due to an increase in $b$
can be offset by an increase in $t_0/t_e$. This degeneracy,
however, does not carry over to the process of low-$l$ multiple
alignment, since the results here do not appear to be sensitive
to the ratio $t_0/t_e$. So we will use the same value ($t_0/t_e=
5\times 10^3$) in every simulation.  By way of interest,
we note that a redshift of $\sim 1100$ (associated with the
surface of last scattering in $\Lambda$CDM) corresponds
to a ratio $t_0/t_e=(1+z)\sim 10^3$ in $R_{\rm h}=ct$.
As is the case in the standard model, the absolute scale is not
known a priori, so the amplitude is also adjustable, e.g.,
by fitting $\delta T_5$ to the data.

  \begin{figure}[h]
   \centering
      \includegraphics[angle=0,width=8.3cm]{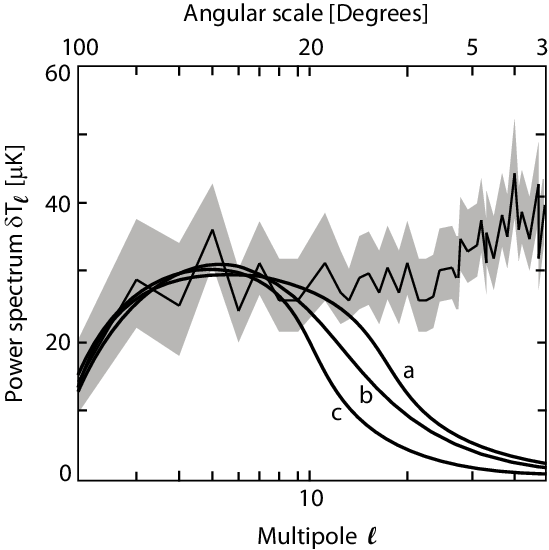}
      \caption{Comparison of theoretical CMB power spectra
       due solely to Sachs-Wolfe-induced fluctuations for
       $N_{SW}=5,000$ and various values of $b$: a) (same
       as figure~1) $b=12$, b) $b=8$, and c) $b=4$.
       The gray region represents the one-$\sigma$ uncertainty.}
\end{figure}

The theoretical curve shown in Figure~1 was calculated
using $b=12$ and $N_{SW}=5,000$, which fits the WMAP
power spectrum quite well for $l<20$, corresponding to
angular scales $>10^\circ$. Not surprisingly, the Sachs-Wolfe
fluctuations produce very little power on scales smaller than
this, where more localized effects (e.g., near the surface
of last scattering) dominate the perturbation growth.
We note, in particular, that the power spectrum predicted
by the $R_{\rm h}=ct$ Universe at $l\rightarrow 1$
agrees with the relative lack of power observed
for the low-$l$ multipoles, a result we had previously
discussed in the context of the CMB angular correlation
function (Melia 2014). Planck has recently confirmed the
surprising WMAP observation of a lack of correlation at
angles $>60^\circ$ (Ade et al. 2013), which does not appear
to be consistent with an inflationary scenario. As we
discussed previously, the downturn in power towards small
values of $l$ in the $R_{\rm h}=ct$ Universe is entirely
due to the maximum angular size $\theta_{\rm max}$ of
fluctuations expected in this cosmology (see Equation~22).

  \begin{figure}[h]
   \centering
      \includegraphics[angle=0,width=8.3cm]{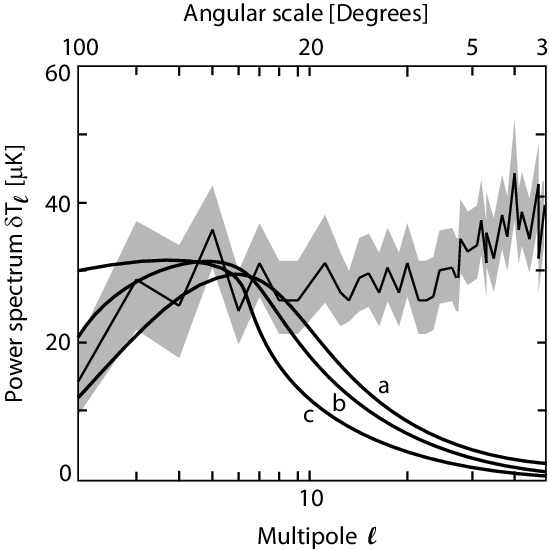}
      \caption{Same as Figure~2, except here for
       $N_{SW}=30,000$. The curves correspond to
       a) $b=12$, b) $b=8$, and c) $b=4$.
       The gray region represents the one-$\sigma$ uncertainty.}
\end{figure}

A comparison of curves in Figures~2--4 demonstrates
how these results depend on $b$ and $N_{SW}$. We see
in Figure~2 that the spectrum below $l\sim 10$ is rather
insensitive to the precise value of $b$. Eventually, when
other small-scale physical processes are included, it may
be possible to more tightly constrain $b$ based on a
comparison of the spectrum with the data at $l\sim 10-30$.
Figure~3 shows that fluctuation numbers $N_{SW}$ much
greater than $5,000$ (in this case, $30,000$) do not produce
a power spectrum matching the CMB's observed features
at $l < 10$. These spectra are either too sharply peaked
at $l\sim 5-6$ or, in the case of c), show too much power
at $l < 5$. And in figure~4, we see that a value of
$N_{SW}$ as low as $1,000$ may work with a relatively
high value of $b$, i.e., $b\sim 12$, corresponding to
curve a), but probably not for smaller values of this
parameter, which seem to produce power spectra
that are too sharply peaked at $l\sim 5$.

  \begin{figure}[h]
   \centering
      \includegraphics[angle=0,width=8.3cm]{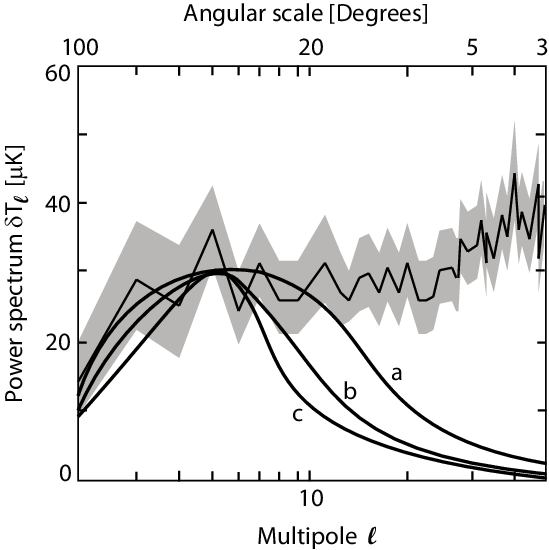}
      \caption{Same as Figure~2, except here for
      $N_{SW}=1,000$. The curves correspond to
      (a) $b=12$, (b) $b=8$, and (c) $b=4$.
      The gray region represents the one-$\sigma$ uncertainty.}
\end{figure}

All in all, this brief survey of the available range
of $N_{SW}$ and $b$ shows that the
number $N_{SW}$ of Sachs-Wolfe fluctuations greater
than $\sim 5,000$ is unlikely to fit the observed CMB
power spectrum at large angles, while a number
$< 1,000$ produces a power distribution too highly
peaked at $l\sim 2-5$. These simulations suggest that
in the $R_{\rm h}=ct$ Universe, $N_{SW}$ is
several thousand, and probably no bigger than
$\sim 5,000$.

\section{The CMB Quadrupole and Octopole Moments}
In Figure~5, we show three simulated renderings of the large-scale
fluctuations in the CMB temperature using values of $b$ ($\sim 3$)
and $t_0/t_e$ ($\sim 5\times 10^3$) indicated by our earlier fits
to the angular correlation function, and $N_{SW}=1,000$. We emphasize
again that none of the effects thought to produce fluctuations on
$< 1^\circ$ scales, such as acoustic oscillations and the various
processes producing secondary signatures after decoupling, are
included in these images (see Melia 2014, and references cited
therein, for a more complete discussion of all the relevant physical
mechanisms). Previous work has shown that these other processes are not
directly relevant to the $l=2$ and $l=3$ multipole moments. The principal
features evident in this figure are due solely to the Sachs-Wolfe effect,
but strictly adhering to the restrictions imposed by Equations~(21) and (26).

\begin{figure}[hp]
{\centerline{\epsscale{0.6} \plotone{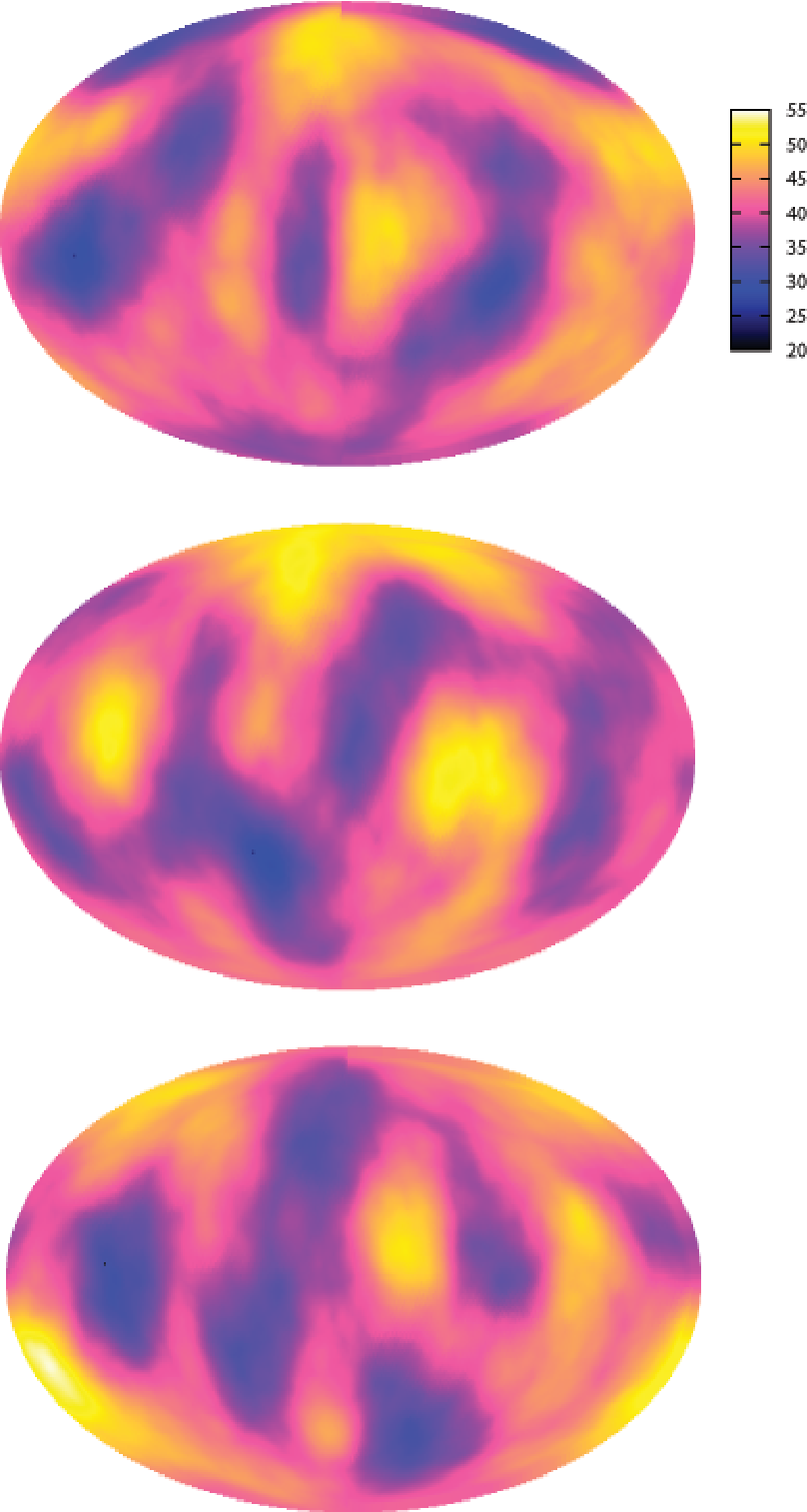} }}
\figcaption{Three simulated renderings of the large-scale
fluctuations in the CMB temperature for the $R_{\rm h}=ct$ Universe.
Here, $t_0/t_e=5\times 10^3$ and $b=3$. Each image
contains $N_{SW}=1,000$ large-size fluctuations. The units on the
color scale are arbitrary.} 
\end{figure}

Before entering into a quantitative statistical analysis of these
simulated all-sky maps, it is quite evident even by eye that the general
features emerging from the $R_{\rm h}=ct$ Universe are reminiscent of those
actually seen in the WMAP data. Note in particular the apparent ``planarity"
of the fluctuations, and the emergence of ``finger-like" darker regions. The
apparent planar-like arrangement of the octopole components was first
noted by de Oliveira et al. (2004), but revisited by many authors since
then. In the analysis of Park et al. (2007), the probability of
observing such a planarity within the context of the standard model is
over $18\%$, and therefore not statistically significant. It is comforting
from the standpoint of the $R_{\rm h}=ct$ Universe that this feature appears
to be present most of the time. The finger-like depressions are more difficult
to quantify, but were noted by Bennett et al. (2011). Again, it is apparent
from these simulations that such features are rather common in the
$R_{\rm h}=ct$ Universe.

Let us now examine how much impact the restricted range
of fluctuation angles ($\theta<\theta_{\rm max}$) has on the distribution
of $\theta_{23}$. We have produced 20,000 simulated all-sky CMB maps for
each assumed value of $N_{SW}$, ranging from 5 to 5,000. As noted earlier,
a sample of these for $N_{SW}=1,000$ is shown in figure~5.

Of course, the number $N_{SW}$ of Sachs-Wolfe-induced
fluctuations cannot be predicted from theory. Only their growth
rate is quantifiable using equations~(11) to (17). For this reason,
we must rely on the outcome not having a strong dependence
on this parameter, which we therefore vary over such a broad
range of values in order to ensure that the statistical properties
associated with the quadrupole and octopole orientations are
not overly influenced by it. The results will show that the
principal reason for the differences between $\Lambda$CDM and
$R_{\rm h}=ct$ is in fact the maximum size of the fluctuations,
set by the gravitational horizon at the time of last scattering
(equations~18-22).

For each synthetic map, we followed the procedure outlined in \S2 above,
using the various techniques described in Appendix A of de Oliveira et al.
(2004) to find the $a_{lm}(\hat{\bf n})$ coefficients in the rotated
frame. From these, we calculated the values of $\theta_{23}$ (between the
quadrupole and octopole moments) and determined their occurrence rate.
The corresponding relative probabilities are shown in Figure~6, for
$N_{SW}=$ 5, 20, 50, and 5,000.

  \begin{figure}[h]
   \centering
      \includegraphics[angle=0,width=16cm]{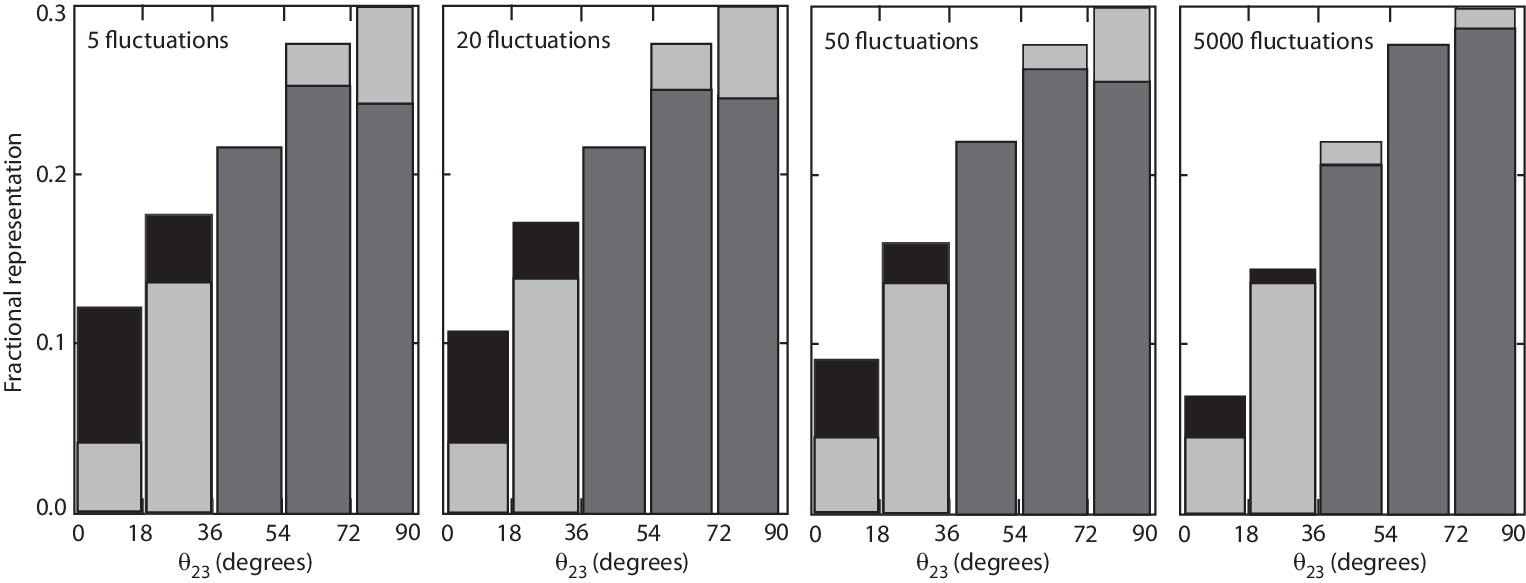}
      \caption{Fractional representation of the angle $\theta_{23}$ between the
CMB quadrupole and octopole moments in 20,000 simulated renderings of the
$R_{\rm h}=ct$ Universe, assuming a total number of 5, 20, 50, and 5,000
large-size fluctuations, respectively (left to right). The light-grey bars show
the fractions in a $\Lambda$CDM universe with fluctuations sampled from
a statistically isotropic, Gaussian random field of zero mean. The black and
dark-grey bars show the corresponding fractions for the $R_{\rm h}=ct$
Universe.}
\end{figure}

Mindful of the conclusions drawn by Park et al. (2007), in which the
observed alignment angle appears to fall within the range
$3.8^\circ<\theta_{23}<18.2^\circ$, we have chosen to
subdivide the results into increments of $18^\circ$, so that the
most likely value of the measured $\theta_{23}$ lies within the first bin.
The probability distribution for a completely random occurrence of the
angle $\theta_{23}$ corresponds to the light-grey bars in these
diagrams, essentially the profile expected in the standard model.
The probability of alignment within the first bin is the aforementioned
value of $4.9\%$, so the observed angle $\theta_{23}$
constitutes a marginally statistically significant anomaly. In contrast,
the probabilities expected for the $R_{\rm h}=ct$ Universe correspond
to the black and dark-grey bars.

Notice that the overall probability distribution depends on how
many fluctuations are included in the simulation. As $N_{SW}$
increases to very large values, we approach the result expected
for $\Lambda$CDM, presumably because this situation is similar
to what one gets with a fluctuation spectrum inflated to very
large scales early in the Universe's history. Absent inflation,
however, the probability distribution is noticeably different,
particularly for the smaller values of $\theta_{23}$. For
$N_{SW}=50$, a value of $\theta_{23}<18^\circ$
is expected to occur about $10\%$ of the time---higher
for smaller fluctuation numbers $N_{SW}$. And for $N_{SW}=5,000$,
this probability is approximately $7\%$. In every case, the
increase in fractional representation at smaller angles is
compensated by the reduced representation at angles
$>50^\circ$ (the dark-grey bars in these diagrams). These
results suggest that an alignment of the quadrupole and
octopole moments to within $\sim 18^\circ$ of each other is
more likely to occur in the $R_{\rm h}=ct$ Universe than in
$\Lambda$CDM, as long as the number of Sachs-Wolfe fluctuations
across the sky is smaller than several thousand, though in both
cosmologies, the probability is always $\leq 10\%$.

\section{Discussion and Conclusions}
It was shown in a detailed analysis of the CMB large-scale anomalies by
Sarkar et al. (2011) that there is no statistically significant correlation
in $\Lambda$CDM between the missing power on large angular scales and the
alignment of the $l=2$ and $l=3$ multipoles. If the CMB anomalies
are not due to astrophysical, instrumental, or data analysis effects,
but instead arise from cosmological influences in the early Universe,
and if we ignore possible biases introduced by the posterior
selection of these two particular features in the CMB,
then the tension between the standard model and the WMAP data is
greater than for each anomaly alone, because their combined statistical
significance could be as small as the product of their individual significances.

As we have noted in the introduction, there are good reasons to believe
that many effects may be responsible for the observed anomalies, so it
is not clear that one should place too much confidence on how these
features impact the models themselves. In particular, one should take
note of the fact that when the contribution of the ISW effect to the
anisotropy pattern is removed from the WMAP map, the statistical
significance of the apparent quadrupole-octopole alignment decreases
considerably (Francis and Peacock 2010; Bennett et al. 2011). If this
is the dominant cause of the alignment, then clearly it has very little
to do with physics in the early Universe.

In this paper, we have addressed the question of whether the
observed alignment may in fact be due to cosmological influences, and
if so, whether its properties may be used to discriminate between
competing models. We have sought to calculate the probability of
seeing this feature in the CMB anisotropies for the
$R_{\rm h}=ct$ Universe. Earlier, we had demonstrated that the angular
correlation function in this cosmology apparently agrees with the WMAP
and {\it Planck} observations without the need to invoke cosmic variance.
As such, the overall probability of seeing no power on large angular scales
and an apparent alignment of the low-$l$ multipoles is due predominantly to
the latter. Though not directly related, these two features of the CMB
nonetheless have a common origin in the $R_{\rm h}=ct$ Universe---the
existence of a maximum angular size $\theta_{\rm max}$ for the large-scale
fluctuations, imposed by the gravitational horizon $R_{\rm h}$ at the time
$t_e$ of last scattering. Our conclusion from this work is that in
$R_{\rm h}=ct$, the simultaneous observation of the missing
large-angle correlations and low-$l$ multipole alignment is likely
at the $7-10\%$ level, depending on what the actual number of
Sachs-Wolfe fluctuations $N_{SW}$ turns out to be. It is
also useful to point out that the increase in the probability of
alignment expected in the $R_{\rm h}=ct$ Universe for $C_i$
and $C_{i+1}$ drops rapidly to levels comparable to those in
$\Lambda$CDM for $i>2$.  For example, the fractional representation 
for $\theta_{34}$ using 5,000 $N_{SW}$ fluctuations (see figure~6) 
is already $\sim5\%$, comparable to the value ($\sim4.9\%$) in 
$\Lambda$CDM. The probability for higher values of $i$ is 
indistinguishable from a completely random occurrence.

Aside from the quantitative aspects of this analysis, a qualitative
comparison between the simulated sky maps shown in Figure~5, and the
real Universe as revealed by WMAP and {\it Planck}, also suggests a
morphological similarity between the two. We noted the appearance of
``finger-like" darkened extensions and the planarity of the octopole
components which, however, are not statistically significant, even in
the standard model. Overall, the weight of evidence---the angular
correlation function, the smaller statistical significance of the
alignment of the quadrupole and octopole moments, and the
morphological similarity between the real and simulated CMB maps---seems
to favor the $R_{\rm h}=ct$ Universe over $\Lambda$CDM, if these
features are all due to the cosmology itself rather than to the many other
possible causes proposed since the observations.

Clearly, there is still work to do. By necessity, our analysis of the
angular correlation function and the low-multipole alignment has
relied on a highly simplified treatment of the fluctuation growth in the
early Universe. It is well known, however, that there are many mechanisms
producing density perturbations, on small and large scales, and there is
a great deal of astrophysics linking these to the actual temperature
variations we see across the sky. Our approach here has merely shown
promise in accounting for the observations. We cannot be completely
certain of the outcome until we have developed a more sophisticated
treatment of the fluctuation growth in the $R_{\rm h}=ct$ Universe,
commensurate with the level of detail already incorporated into the
standard model.

\vskip-0.2in
\acknowledgments
I am grateful to the anonymous referee for his careful review of
this paper and for recommending several important improvements to its
content. This research was partially supported by ONR grant N00014-09-C-0032 
at the University of Arizona, and by a Miegunyah Fellowship at the 
University of Melbourne. I am particularly grateful to Amherst College 
for its support through a John Woodruff Simpson Lectureship.

\end{document}